\documentclass[pre,twocolumn,amssymb,showpacs]{revtex4}

\usepackage{graphicx}
\usepackage{color}
\usepackage{epsfig}
\usepackage{latexsym}
\usepackage{bm}
\usepackage{ulem}

\usepackage{color}

\begin{document}

\renewcommand{\ni}{{\noindent}}
\newcommand{\dprime}{{\prime\prime}}
\newcommand{\be}{\begin{equation}}
\newcommand{\ee}{\end{equation}}
\newcommand{\bea}{\begin{eqnarray}}
\newcommand{\eea}{\end{eqnarray}}
\newcommand{\nn}{\nonumber}
\newcommand{\bk}{{\bf k}}
\newcommand{\bQ}{{\bf Q}}
\newcommand{\q}{{\bf q}}
\newcommand{\s}{{\bf s}}
\newcommand{\bN}{{\bf \nabla}}
\newcommand{\bA}{{\bf A}}
\newcommand{\bE}{{\bf E}}
\newcommand{\bj}{{\bf j}}
\newcommand{\bJ}{{\bf J}}
\newcommand{\bs}{{\bf v}_s}
\newcommand{\bn}{{\bf v}_n}
\newcommand{\bv}{{\bf v}}
\newcommand{\la}{\langle}
\newcommand{\ra}{\rangle}
\newcommand{\dg}{\dagger}
\newcommand{\br}{{\bf{r}}}
\newcommand{\brp}{{\bf{r}^\prime}}
\newcommand{\bq}{{\bf{q}}}
\newcommand{\hx}{\hat{\bf x}}
\newcommand{\hy}{\hat{\bf y}}
\newcommand{\bS}{{\bf S}}
\newcommand{\cU}{{\cal U}}
\newcommand{\cD}{{\cal D}}
\newcommand{\bR}{{\bf R}}
\newcommand{\pll}{\parallel}
\newcommand{\sumr}{\sum_{\vr}}
\newcommand{\cP}{{\cal P}}
\newcommand{\cQ}{{\cal Q}}
\newcommand{\cS}{{\cal S}}
\newcommand{\ua}{\uparrow}
\newcommand{\da}{\downarrow}
\newcommand{\red}{\textcolor {red}}
\newcommand{\blu}{\textcolor {blue}}
\newcommand{\1}{{\oldstylenums{1}}}
\newcommand{\2}{{\oldstylenums{2}}}
\newcommand{\mDelta}{\varepsilon}
\newcommand{\m}{\tilde m}
\def\lsim {\protect \raisebox{-0.75ex}[-1.5ex]{$\;\stackrel{<}{\sim}\;$}}
\def\gsim {\protect \raisebox{-0.75ex}[-1.5ex]{$\;\stackrel{>}{\sim}\;$}}
\def\lsimeq {\protect \raisebox{-0.75ex}[-1.5ex]{$\;\stackrel{<}{\simeq}\;$}}
\def\gsimeq {\protect \raisebox{-0.75ex}[-1.5ex]{$\;\stackrel{>}{\simeq}\;$}}

\title{Additivity, density fluctuations, and nonequilibrium thermodynamics for active Brownian particles}

\author{Subhadip Chakraborti} 
\email{subhodip.chakraborti@bose.res.in}
\author{Shradha Mishra}
\email{smishra.phy@itbhu.ac.in} 
\author{Punyabrata Pradhan}
\email{punyabrata.pradhan@bose.res.in}

\affiliation{ Department of Theoretical Sciences, S. N. Bose National Centre for Basic Sciences, Block - JD, Sector - III, Salt Lake, Kolkata 700106, India }

\begin{abstract}

\noindent{   
Using an additivity property, we study particle-number fluctuations in a system of interacting self-propelled particles, called active Brownian particles (ABPs), which consists of repulsive disks with random self-propulsion velocities. From a fluctuation-response relation - a direct consequence of additivity, we formulate a  thermodynamic theory which captures the previously observed features of nonequilibrium phase transition in the ABPs from a homogeneous fluid phase to an inhomogeneous phase of coexisting gas and liquid. We substantiate the predictions of additivity by analytically calculating the subsystem particle-number distributions in the homogeneous fluid phase away from criticality where analytically obtained distributions are compatible with simulations in the ABPs.     
}

\typeout{polish abstract}

\end{abstract}


\maketitle

\section{Introduction}

The problem of characterization of driven many-particle systems, having a steady state, has drawn much attention in the past decades \cite{Gallavotti, Derrida}. The problem is however quite hard to tackle, mainly due to that the steady-state probability weights of microscopic configurations in a driven system, unlike in equilibrium, are not described by the Boltzmann distribution. Indeed, in most cases, these weights are not known. A simple characterization of the steady-state systems in general would be certainly desirable and, to this end, various attempts have been made in the past \cite{Oono, Bertini}. Recently, a formulation based on equilibriumlike additivity property provides a framework \cite{Eyink1996, Bertin_PRL2006, PRL2010}, which helps one to describe a broad class of nonequilibrium steady states through fluctuations of a conserved quantity, e.g., mass or particle-number \cite{Chatterjee_PRL2014, Chatterjee_PRE, Das_PRE2015}. Here, we address the question whether an additivity property can be used to obtain large deviation probability for the particle-number or the density fluctuations, the central object in a statistical mechanics theory, in systems of self-propelled particles.

Self-propelled particles (SPPs), also called active matters, are prevalent in nature - in living systems, e.g., bacterial colony \cite{Jacob_AdvPhys2000}, fish school \cite{Hubbard}, flocks of birds \cite{Feder}, insect swarm \cite{Rauch} as well as in nonliving systems, e.g., photoactivated or chemically powered colloids, thermophoretic Janus particles \cite{Nonliving}, etc. They have been realized in experiments \cite{Experiments} and intensively studied through simulations and theories \cite{Mishra_PRE2010, Peruani_PRL2011, Bertin_PRL2012, Fily_Marchetti_PRL2012, Farrell_PRL2012, Speck_PRL2012, Baskaran_PRL2013, Speck_EPL2013, Bertin_PRL2014}; for reviews, see \cite{Marchetti_RMP2013, Cates_condmat2015}. These particles propel themselves by converting chemical energy to mechanical one, which is continually dissipated to the medium. The steady flow of energy keeps the system out of equilibrium and a novel nonequilibrium steady state (NESS) emerges. Such a nonequilibrium steady state manifests itself by exhibiting rich collective phenomena, e.g., self-assemblies and pattern formations, otherwise impossible in equilibrium.

Recently, we have seen  a surge of activities, in search of a suitable statistical mechanics framework which could describe macroscopic properties of the self-propelled particles in terms of an intensive thermodynamic variable, such as a chemical potential \cite{Cates_PRL2008, Cates_PNAS2010, Cates_EPL2013, Cates_PRL2013, Speck_PRL2014, Cates_NatCom2014, Peruani_JSP2015, Cates_condmat2015, Joanny_condmat2015}, pressure \cite{Brady_PRL2014, Mallory_PRE2014, Tailleur_PRL2015, Tailleur_Nature2015, Marchetti_SM2014, Gompper} or an effective temperature \cite{Berthier_condmat2015, Marconi_condmat2015}. However, a complete framework still remains elusive. We propose here a general thermodynamic principle, called additivity, which could enable us to unify fascinatingly 
broad-ranging phenomena in the systems of self-propelled particles under a unique nonequilibrium thermodynamic theory, directly connecting microscopic fluctuations to the macroscopic properties in the system.

In this paper, using an additivity property, we formulate a thermodynamic theory for a particular class of self-propelled particles, called active Brownian particles (ABPs), consisting of repulsive disks in two dimensions and which have random self-propulsion velocities. We demonstrate, in the regime of homogeneous phase, that additivity property leads to subsystem particle-number distribution $P_{\cal V}({\cal N})$, the probability that a subsystem of volume ${\cal V}$ has ${\cal N}$ number of particles. The logarithm of the probability $P_{\cal V}({\cal N})$, or the large deviation function (LDF) - analogous to equilibrium free energy, governs the density fluctuations and thus can immediately connect to the standard statistical mechanics framework. We validate the predictions of the theory regarding density fluctuations by analytically calculating the subsystem particle-number distribution $P_{\cal V}({\cal N})$ in the homogeneous fluid phase in the active Brownian particles and comparing them with simulations.

The crucial ingredient of this theory is a nonequilibrium fluctuation-response relation (FR) between compressibility and number-fluctuation or variance $\sigma^2_{\cal V} = \langle {\cal N}^2 \rangle - \langle {\cal N} \rangle^2$ (see Eq. \ref{FR1}), which is a direct consequence of additivity. Provided the functional dependence of the variance $\sigma^2_{\cal V}(\rho)$ on the particle-number density $\rho$, we provide a prescription of how,  using additivity, one can calculate the distribution function $P_{\cal V}({\cal N})$.

To illustrate the formalism, we first calculate, within a linearized fluctuating hydrodynamics of the ABPs, the variance $\sigma^2_{\cal V}(\rho)$ of particle-number in a subsystem of volume ${\cal V}$ as a function of density $\rho$. Then, we use the standard large deviation methods to obtain the large deviation function, or a nonequilibrium free energy density function $f(\rho, Pe)$, and a chemical potential $\mu(\rho, Pe)$, as a function of number density $\rho$ and activity parameter Peclet number $Pe$. Determination of chemical potential leads to a nonequilibrium equation of state - akin to the equilibrium Van der Waals one. Beyond a critical activity, compressibility $d\rho/d\mu$ becomes negative in a particular density interval, leading to nonmonotonic $\mu$ as a function of $\rho$ and hence phase coexistence. In special limits, our theory captures various previous results, e.g., those based on the concept of motility induced phase separation (MIPS) \cite{Cates_PRL2008, Cates_EPL2013, Cates_PRL2013}, indicating the formulation here is indeed consistent with the past studies. Moreover, our analysis suggests that, on a mean-field level, a broad class of self-propelled particles belong to Ising universality.

The organization of the paper is as follows. In section \ref{sec-additivity}, we discuss additivity and show how subsystem particle-number distribution can be calculated solely from the knowledge of variance of subsystem particle-number as a function of number density. In section \ref{Model-ABP}, we define the model of active Brownian particles and discuss the corresponding fluctuating hydrodynamics. In section \ref{Variance}, we calculate, within linearized hydrodynamics, variance of subsystem particle number as a function of density (section \ref{Variance-linear}) and then characterize noise strengths in the hydrodynamic equations (section \ref{noise-strength}). In section \ref{noneq_therm}, using the functional dependence of the variance on density together with additivity, we formulate a thermodynamic theory of active Brownian particles and substantiate the theory by explicitly calculating subsystem particle-number distributions in homogeneous fluid phase of active Brownian particles. Finally, we summarize in section \ref{Summary}.

\section{Additivity and subsystem particle-number distribution} 
\label{sec-additivity}

In this section, we discuss an additivity property, which systems having a {\it finite} correlation length are expected to possess, {\it irrespective of} whether the systems are {\it in} or {\it out} of equilibrium \cite{Eyink1996, Bertin_PRL2006, PRL2010}. Recently, additivity has been used in nonequilibrium mass-transport processes for calculating mass distributions and characterizing macroscopic properties in terms of equilibriumlike thermodynamic potentials \cite{Chatterjee_PRL2014, Chatterjee_PRE}.  Below, we discuss how additivity can be used to calculate subsystem particle-number distribution.

First, let us discuss what additivity means in the context of 
particle-number or density fluctuations in a system. Let us consider $N$ interacting particles in volume $V$ where the total number of particles $N$ is conserved. We divide the system in $\nu=V/{\cal V}$ number of identical subsystems, each having volume ${\cal V}$, and ask what could be the form of the joint probability distributions for the subsystem particle-numbers $\{ {\cal N}_i \} \equiv \{ {\cal N}_1, {\cal N}_2, \dots, {\cal N}_{\nu} \}$. Provided that the subsystem size is much larger than spatial correlation length $\xi$, ${\cal V}^{1/d} \gg \xi$ in $d$ dimensions, additivity implies that the subsystems are statistically almost independent and therefore, to a very good approximation, the joint subsystem particle-number distribution can be written in a product form \cite{Eyink1996, Bertin_PRL2006, PRL2010},
\be {\cal
P}[\{{\cal N}_i \}] \simeq \frac{\prod_{k=1}^{\nu} W_{\cal V}({\cal N}_k)}{Z(N, V)} \delta\left( \sum_k {\cal N}_k - N \right), \label{additivity1} 
\ee 
in the thermodynamic limit of $N, V \rightarrow \infty$ with density $\rho=N/V$ fixed. In Eq. \ref{additivity1}, $W_{\cal V}({\cal N}_k)$ is an unknown weight factor, which depends on the subsystem particle-number, and will be determined later. The normalization constant, or the partition sum, $Z(N,V)$ in Eq. \ref{additivity1} can be written as
\be
Z(N, V) = \sum_{\{{\cal N}_i \}} \left[ \prod_k W_{\cal V}({\cal N}_k) \right] \delta \left( \sum_k {\cal N}_k - N \right).
\ee 
In other words, the property that the joint subsystem 
particle-number distribution ${\cal P} [\{ {\cal N}_i \}]$ for a system can be approximately written as a product  of individual subsystem weight factors $W_{\cal V}({\cal N}_k)$ (i.e., subsystems are statistically almost independent) is called additivity.

In equilibrium, there is a well-defined thermodynamic prescription, which helps us to calculate the weight factor $W_{\cal V}({\cal N}_k)$, i.e., $W_{\cal V}({\cal N}_k)$ can in principle be obtained from the Boltzmann distribution. However, there is no such prescription in nonequilibrium. In fact, in nonequilibrium, the difficulty arises precisely here because, in most cases, the microscopic weights of the configurations in the steady state are a priori unknown. At this scenario, additivity, which originates from the simple physical consideration of statistical independence on the coarse-grained level of large subsystems, could help us to bypass the difficulty. As demonstrated recently in \cite{Chatterjee_PRL2014, Chatterjee_PRE}, to characterize fluctuation properties on a coarse-grained level, one may not actually be required to obtain the steady-state weights of all microscopic configurations. In fact, obtaining coarse-grained probability weights on a large scale (much larger than the microscopic correlation length scale) would suffice to characterize the macroscopic properties of the system, provided that additivity as in Eq. \ref{additivity1} holds.

It is important to note that the weight factor $W_{\cal V}({\cal N}_k)$ depends only on the subsystem particle-number ${\cal N}_k$ and subsystem volume ${\cal V}$. Now, provided that Eq. \ref{additivity1} holds, probability distribution function $P_{\cal V}({\cal N}) \equiv {\rm Prob}[{\cal N}_k = {\cal N}]$ for large ${\cal V}$ can be written as \cite{Eyink1996, Bertin_PRL2006, PRL2010, Chatterjee_PRL2014},
\bea
P_{\cal V}({\cal N}) &\simeq& W_{\cal V}({\cal N}) \frac{Z(N-{\cal N}, V - {\cal V})}{Z(N, V)} 
\nonumber \\
&=& \frac{W_{\cal V}({\cal N}) e^{\mu(\rho) {\cal N}} }{\cal Z},
\label{Pn1}
\eea
where $\mu(\rho)$ is a nonequilibrium chemical potential,
\be
\mu(\rho) = \frac{df}{d\rho},
\ee 
$f(\rho)$ is a nonequilibrium free energy density function with $Z(N,V) \simeq \exp[-Vf(\rho)]$ and ${\cal Z} = \sum_{\cal N} W_{\cal V}({\cal N}) \exp(\mu {\cal N})$ is the normalization constant. Importantly, free energy density function $f(\rho)$, or equivalently the large deviation function (LDF) which controls the density fluctuations, and chemical potential $\mu(\rho)$ can now be obtained from a fluctuation-response relation (FR) between compressibility and fluctuation \cite{Chatterjee_PRL2014, Eyink1996, Bertin_PRL2006, PRL2010}, 
\be
\frac{d \rho}{d \mu} = \sigma^2(\rho),
\label{FR1}
\ee
where 
\be 
\sigma^2(\rho) = \lim_{{\cal V} \rightarrow \infty}  \frac{(\langle {\cal N}^2 \rangle - \langle {\cal N} \rangle^2)}{\cal V},
\ee
the scaled variance of subsystem particle-number ${\cal N}$. The above fluctuation-response relation is analogous to the equilibrium fluctuation-dissipation theorem and follows directly from Eq. \ref{additivity1}; for details, see Appendix A. The explicit expression of chemical potential $\mu(\rho)$ and free energy density function $f(\rho)$ are obtained by integrating Eq. \ref{FR1} w.r.t. density $\rho$, 
\be
\mu(\rho) = \int \frac{1}{\sigma^2(\rho)} d\rho + c_1,
\ee 
and, upon further integration, 
\be 
f(\rho) = \int \mu(\rho) d\rho + c_2,
\ee 
where $c_1$ and $c_2$ are arbitrary integration constants.

We now show, following \cite{Das_PRE2015, Touchette}, that the weight factor $W_{\cal V}({\cal N})$ and, consequently, the particle-number distribution can indeed be calculated using the above chemical potential and free energy function. First we write Laplace transform (discrete) of the partition sum $Z(N, V)$) as
\bea
&& \tilde Z(s,V)  =  \sum_{N=0}^{\infty} e^{-s N} Z(N, V) 
\nonumber \\
&=& \sum_{N=0}^{\infty} e^{-s N} \sum_{\{ {\cal N}_k \} } \left[ \prod_{k=1}^{\nu={\cal V}/V}  W_{\cal V}({\cal N}_k) \right] 
\delta \left(\sum_k {\cal N}_k - N \right)
\nonumber \\
&=& \prod_{k=1}^{\nu} \left[ \sum_{{\cal N}_k=0}^{\infty} e^{-s {\cal N}_k} W_{\cal V}({\cal N}_k) \right] = \left[ \tilde{W_{\cal V}}(s) \right]^\nu, \nonumber
\eea
where Laplace transform of the weight factor $W_{\cal V}({\cal N})$ is written as $\tilde{W_{\cal V}} (s) = \sum_{{\cal N}=0}^{\infty} e^{-s {\cal N}} W_{\cal V}({\cal N}).$ Now approximating $\tilde{Z}(s,V) = \sum_{N=0}^{\infty} e^{-s N} Z(N, V) \simeq \int_0^{\infty} dN Z(N, V) e^{-s N}$ where we replace the sum by an integral and then, using $Z(N,V) \simeq \exp[-Vf(\rho)]$ (by definition), we get
$$
e^{-\nu h_{\cal V}(s)} \equiv V \int e^{-V[f(\rho)+ s \rho]} d\rho \simeq [\tilde{W}_{\cal V}(s)]^\nu.
\label{Zs1}
$$
where the function $h_{\cal V}(s)$ is obtained from Legendre transform of free energy density function,
\be
h_{\cal V}(s) = {\cal V} [{\rm \bf inf}_\rho \{ f(\rho)+s \rho\}],
\ee
The weight factor $W_{\cal V}({\cal N})$ can, in principle, be calculated by evaluating the following integral on the complex $s$-plane along a suitably chosen contour $C$: $W_{\cal V}({\cal N}) = 1/(2 \pi i) \int_C \exp[-h_{\cal V}(s)+{\cal N}s] ds$. Although, for finite ${\cal V}$, the explicit calculation of the weight factor may be difficult, the calculation, for large subsystem sizes ${\cal V} \gg \xi$, simplifies as the function $-(1/{\cal V}) \ln W_{\cal V} ({\cal N})$ is related to $h_{\cal V}(s)/{\cal V}$ by Legendre transformation \cite{Touchette}. Therefore, in the leading order of ${\cal N} \gg 1$ and ${\cal V} \gg \xi^d$, the function $-(1/{\cal V}) \ln W_{\cal V} ({\cal N})$ is nothing but the free energy density function $f({\cal N}/{\cal V})$, implying
\be
W_{\cal V}({\cal N}) \simeq \exp[-{\cal V}f({\cal N}/{\cal V})].
\ee
Consequently, the subsystem particle-number distribution can be written as
\be 
P_{\cal V}({\cal N}) \simeq \frac{e^{[-{\cal V}f({\cal N}/{\cal V}) + \mu(\rho) {\cal N}]}}{\cal Z},
\label{P_N_additivity}
\ee
where ${\cal Z}(\mu, V)$ is the normalization constant.

\section{Model of Active Brownian particles and fluctuating Hydrodynamics}
\label{Model-ABP}

To illustrate our theory in a particular model system of 
self-propelled particles, we study fluctuations in a system of interacting active Brownian particles (ABPs) in two dimensions (2D). We consider $N$ particles in a $2D$ periodic box of size $V = L \times L$. At time $t$, the system is specified by position ${\bf R}_i(t)$ and self-propulsion direction $\phi_i(t)$ of $i$th particle with $i= 1, \dots, N$. The system evolves in time through the following over-damped Langevin equations \cite{Baskaran_PRL2013}, for the positions $\{{\bf R}_i(t)\}$
$$
{\bf \dot{R}}_i = -\beta D_0 {\bf F}_i + v_0 {\bf u}_i + \sqrt{2 D_0} \vec{\eta}^T_i
$$ 
and for the orientations $\{ \phi_i(t) \}$ of the velocity vectors
$$
\dot{\phi}_i = \sqrt{2 D_r} \eta^R_i,
$$
where $\beta=1/k_BT$ inverse temperature (we set $\beta=1$), force on $i$th particle ${\bf F}_i = \sum_{j \ne i} \nabla U(|{\bf R}_i - {\bf R}_j|)$, the WCA interaction potential $U(r)=4 \epsilon[(a/r)^{12} - (a/r)^6]+\epsilon$ if $r < 2^{1/6}a$ and zero otherwise, $\epsilon=\beta^{-1}$, $a$ diameter of the particles, $v_0$ self-propulsion speed, ${\bf u}_i \equiv \{u_{ix}, u_{iy} \}=\{\cos \phi_i, \sin \phi_i\}$ unit vector along instantaneous self-propulsion direction, $D_0$ and $D_r$ translational and rotational diffusion constant, respectively, and the $\eta$'s Gaussian white noises with $\langle \eta_i \rangle=0$ and $\langle \eta_i(t) \eta_j(t') \rangle = \delta_{ij} \delta(t-t')$.


To analytically study particle-number fluctuations in the active Brownian particles, we resort to a fluctuating hydrodynamic description, representing the system on a coarse-grained level.
The following hydrodynamic equations, {\it without} the noise terms, for a density field $\rho({\bf r}, t)$ and a polarization density field ${\bf p}({\bf r},t)$ has been previously obtained and studied for the active Brownian particles \cite{Speck_EPL2013, Speck_PRL2014, Cates_condmat2015},
\bea
\label{Hydro1}
\partial_t \rho &=& - {\bf \nabla}.\left[ v(\rho) {\bf p} - D(\rho) {\bf \nabla }\rho+ {\bf f}_d \right],
\\
\label{Hydro2}
\partial_t{\bf{p}} &=& -D_r{\bf{p}}-\frac{1}{2}{\bf{\nabla}}\left(v \rho \right)+ K\nabla^2{\bf{p}}+ {\bf f}_p,
\eea
where $\rho({\bf{r}},t) = \sum_{i} \delta ({\bf r}-{\bf R}_i(t))$ and ${\bf p}({\bf r},t)=\rho ({\bf r},t){\bf P}({\bf{r}},t)=\sum_{i} \delta ({\bf r}-{\bf R}_i(t)) {\bf u}_i(t)$ are coarse-grained number and polarization densities, respectively, at position ${\bf r}$ and time $t$ with ${\bf R}_i(t)$ and ${\bf u}_i(t)$ being position and velocity-direction of the $i$th particle respectively, $D(\rho)$ bulk diffusion constant, $v(\rho)$ bulk velocity, ${\bf f}_d$ and ${\bf f}_p$ Gaussian noises specified below. Note that, to study fluctuations, we have added the noise terms ${\bf f}_d$ and ${\bf f}_p$ \cite{Fily_Marchetti_PRL2012} - Gaussian multiplicative noises with zero mean and correlations $\langle f_{d \nu}({\bf r},t)f_{d \nu'}({\bf r}^\prime, t^\prime) \rangle = 2 \Delta_d(\rho) \delta_{\nu \nu'} \delta({\bf r} - {\bf r}^\prime) \delta(t - t^\prime)$ and $\langle f_{p \nu}({\bf r},t) f_{p \nu'}({\bf r}^\prime, t^\prime) \rangle = 2 \Delta_p(\rho) \delta_{\nu \nu'} \delta({\bf r} - {\bf r}^\prime) \delta(t - t^\prime)$, with $\nu, \nu'=1,2$ denoting Cartesian components. The strengths of the noise terms are not previously known and are characterized later (see section \ref{noise-strength}).

When $v(\rho) \ne 0$, the steady-state probability functional ${\cal P}[\{\rho({\bf r}), {\bf p}({\bf r}) \}]$ \cite{Gardiner} {\it neither} has the Boltzmann distribution for the effective probability ${\cal P}_d[\{\rho({\bf r}) \}] \propto \exp[-\int f[\rho({\bf r})] d^2 {\bf r} ]$ for density, {\it nor} it is in general known; only in special cases, a strictly local free energy functional $f[\rho({\bf r})]$ can be obtained \cite{Cates_PRL2008, Cates_EPL2013, Peruani_JSP2015}. However, additivity in Eq. \ref{additivity1} requires {\it neither} the existence of any Boltzmann-like distribution {\it nor} any prior knowledge of the full steady-state structure; it requires only the existence of a finite correlation length $\xi(\rho)$ (see the relevant length scales $\xi_{0,1,2}$ defined in Eqs. \ref{S1} and \ref{S2}). When $v=0$, ${\cal P}_d[\{\rho({\bf r}) \}]$ can be exactly calculated; for details, see Appendix E.

\section{Variance of Subsystem particle-number} 
\label{Variance}

\subsection{Calculation of variance using linearized hydrodynamics}
\label{Variance-linear}

As discussed in section \ref{sec-additivity}, using the 
fluctuation-response relation Eq. \ref{FR1}, subsystem 
particle-number distribution $P_{\cal V}({\cal N})$ for large ${\cal V}$ can be determined solely from the variance of particle number, which requires knowledge of only two-point correlation function $c({\bf r})= \langle \rho(0) \rho({\bf r}) \rangle - \langle \rho({\bf r}) \rangle^2$. To this end, we transform the variable $\theta({\bf r}, t)=\nabla.{\bf p}$ in Eq. \ref{Hydro2} and, using the standard linear analysis, expand the nonlinear terms in Eqs. \ref{Hydro1} and \ref{Hydro2} upto linear order of $\delta \rho$ and $\delta {\bf p}$, where $\delta \rho = \rho - \rho_0$, $\delta {\bf p} = {\bf p} - {\bf p}_0$, $\delta \theta = \nabla. (\delta {\bf p})$ with ${\rho}_0$ and ${\bf p}_0=0$ average density and polarization fields respectively. Using Fourier transform of $\delta \rho({\bf r}, t)$ and $\delta \theta ({\bf r}, t)$ in the linearlized hydrodynamic equations,
\begin{eqnarray}
\label{eq5}
\delta\tilde{\rho}({\bf{q}},\omega)&=&\int_{\bf{r}}\int_t e^{-i{\bf{q}}.{\bf{r}}} e^{-i\omega t}~\delta \rho({\bf{r}},t) d{\bf r} dt,
\\
\label{eq6}
\delta\tilde{\theta}({\bf{q}},\omega)&=&\int_{\bf{r}}\int_t e^{-i{\bf{q}}.{\bf{r}}} e^{-i\omega t}~\delta \theta({\bf{r}},t) d{\bf r} dt,
\end{eqnarray}
and proceeding along the lines of Ref. \cite{Fily_Marchetti_PRL2012}, we obtain static structure factor $$S({\bf q}) =\frac{1}{2 \pi} \int_{-\infty}^\infty \langle |\delta \tilde{\rho}({\bf q},\omega)|^2 \rangle d\omega \equiv S_1(q) + S_2(q),$$ where
\bea
S_1(q) &=& \frac{V \Delta_d (\Sigma_1)^2}{{\cal D} \Sigma_0 \Sigma_2}  + \frac{V \Delta_d q^2 }{D_r \Sigma_2}, 
\label{S1}
\\  
S_2(q) &=& \frac{V \Delta_p v^2}{D_r^2 {\cal D} \Sigma_0 \Sigma_2},
\label{S2}
\eea
an effective diffusivity ${\mathcal{D}}(\rho) = D + v \alpha / D_r$, $\alpha(\rho) = (v + \rho dv/d\rho)/2$, $\Sigma_{0,1,2}(q)=(1 + q^2 \xi_{0,1,2}^2)$ and correlation lengths $\xi_0(\rho) = \sqrt{D K/D_r {\cal D}}$, $\xi_1=\sqrt{K/D_r}$ and $\xi_2=\sqrt{(D+K)/D_r}$; for details, see Appendix B. Now the variance $\sigma_{\cal V}^2(\rho) = \langle {\cal N}^2 \rangle - \langle {\cal N} \rangle^2$ of particle-number ${\cal N} = \int_{\cal V} \rho({\bf r}) d^2 {\bf r}$ in a subvolume ${\cal V}$ can be written as integrated correlations, $\sigma_{\cal V}^2(\rho)= \int_{\cal V} c({\bf r}) d^2 {\bf r} = S({\bf q}=0)$. By defining a scaled variance $\sigma^2(\rho) = \sigma_{\cal V}^2/{\cal V}$, we finally obtain the variance, albeit within an approximate linearized analysis,  
\begin{eqnarray}
\label{sigma1}
\sigma^2(\rho) =  \left[ \frac{\Delta_d}{\mathcal{D}}  + \frac{\Delta_p v^2}{D_r^2\mathcal{D}} \right], 
\end{eqnarray}
which can be related to compressibility $(d \rho/ d\mu)$ through the fluctuation-response relation Eq. \ref{FR1}. The above linear analysis, though approximate, is expected to be valid in the regime of small fluctuations, i.e., far away from criticality. A similar expression for structure factor was previously obtained in \cite{Fily_Marchetti_PRL2012}, though without the part $S_1({\bf q})$ and without any characterization of the noise strengths $\Delta_d$ and $\Delta_p$. The functional dependence of the noise strengths $\Delta_d$ and $\Delta_p$ on density $\rho$ and self-propulsion speed $v_0$ will be determined later in \ref{noise-strength}.  Note that, in r.h.s. of Eq. \ref{sigma1}, effective diffusivity ${\cal D}(\rho)$ appears in denominators of both the terms, which could vanish  for suitable parameter values; consequently, both $S_1({\bf q}=0)$ and $S_2({\bf q}=0)$ can separately diverge.

There are two interesting limiting cases of Eq. \ref{sigma1}, which  consistently capture various previous results obtained in the context of motility induced phase separation (MIPS) in self-propelled particles. 
\\
{\it Case I.}  To see that $S_1({\bf q})$ in Eq. \ref{sigma1} can have nontrivial effects, we consider the case when the polarization noise vanishes, $\Delta_p=0$. In that case, the above linear analysis implies that $S_1(0)$ diverges at a critical density, for any $\Delta_d$, whenever ${\cal D}=0$ (${\cal D} <0$ corresponds phase coexistence) depending on the functional form of $v(\rho)$. This explains why, in the quasistatic case of ${\bf p}$ where $\partial_t {\bf p}=0$, $K=0$ and $\Delta_p=0$ in Eq. \ref{Hydro2}, the variance $\sigma^2(\rho) \simeq [ {1}/{\rho} + ({1}/{v}) ({dv}/{d\rho}) ]^{-1}$ obtained from Eq. \ref{sigma1} by choosing $\Delta_d = (D +v^2/2D_r) \rho$ as in \cite{Cates_PRL2008} and assuming $D \ll v^2/2D_r$ (large velocity regime), can be diverging (for details, see Appendix D). Because, chemical potential $\mu(\rho) = \ln (\rho v) + c_1$, obtained using the fluctuation-response relation Eq. \ref{FR1}, has a singularity at the critical point where $dv/d\rho=-v/\rho$ and consequently compressibility $d\rho/d\mu=(d^2f/d\rho^2)^{-1}$ diverges; the spinodal line is provided by the condition $d^2f/d\rho^2 < 0$, which is consistent with the previous observations in various systems of self-propelled particles \cite{Cates_PRL2008, Peruani_JSP2015, Tailleur_PRL2015}.  
\\
{\it Case II.} On the other hand, in the absence of density noise, $\Delta_d=0$ \cite{Fily_Marchetti_PRL2012}, only the second term in the r.h.s. of Eq. \ref{sigma1} contributes to $\sigma^2(\rho) = ( {2 \Delta_p^{0}}/{D_r} ) \left[ {D}/{\rho v^2} + \{ {1}/{\rho} + ({1}/{v}) ({dv}/{d\rho}) \} \right]^{-1}$ as, from the central limit theorem (CLT), the polarization noise strength $\Delta_p \simeq \Delta_p^{0} \rho$ is proportional to the number of particles present in unit volume (for details, see Appendix C). Integrating fluctuation-response relation Eq. \ref{FR1}, we obtain $\mu(\rho) = \ln (\rho v) + \psi(\rho) + c_1$ and $f(\rho) = \int \mu(\rho) d\rho = \rho (\ln \rho -1) + \int^{\rho} [\ln v(\rho) + \psi(\rho)] d\rho + c_1 \rho + c_2$ where $\psi'(\rho)=D/\rho v^2$ and $c_1$ and $c_2$ arbitrary constants of integration. Indeed, the above expressions of $\mu(\rho)$ and $f(\rho)$ are quite similar to those obtained for the MIPS in the self-propelled particles (see Case I).

\subsection{Calculation of noise-strengths and fluctuations}
\label{noise-strength}

The main difficulty to relate fluctuating hydrodynamic equations \ref{Hydro1} and \ref{Hydro2} to the microscopic model of active Brownian particles lies in the fact that the noise strengths $\Delta_d$ and $\Delta_p$, bulk diffusion constant $D$ and bulk velocity $v$ could depend on density $\rho$, self-propulsion velocity $v_0$ (even function of $v_0$) and possibly on the norm $|{\bf p}|$, but their functional forms are not explicitly known. In fact, a systematic derivation of the noise strengths from a microscopic dynamics is a difficult problem and, so far, has not achieved for the active Brownian particles.

In this section, we characterize strengths of the noises in the hydrodynamic equations \ref{Hydro1} and \ref{Hydro2}, in the leading order of self-propulsion velocity $v_0$, i.e., when activity is low.  
To this end, we resort to a near-equilibrium analysis, which, we see later in simulations, however holds surprisingly well even far away from equilibrium where self-propulsion velocity, or the activity, is quite large. We first note that equilibrium compressibility of two-dimensional hard-disk fluid, known through virial coefficients \cite{review_virial}, has an approximate analytic form \cite{Hoste},
\be
\left[ \sigma^2(\rho)\right]_{v_0=0} = \frac{\Delta_d}{D} \simeq \rho \left( 1 - \frac{\rho}{\rho_m} \right)^2,
\label{sigma_eq}
\ee  
where $\rho_m \approx 1.15$ close-packing density. Now, we expand $\Delta_d$ in the leading order of self-propulsion velocity $v_0$,
\be 
\Delta_d(\rho, v_0)  \simeq  (\Delta_d^0 + \Delta_{d}^1 v_0^2) \rho \left( 1- \frac{\rho}{\rho_m} \right)^2
\ee 
and write $D \simeq D_0$ where $\Delta_d^0$, $\Delta_{d}^1$ and $D_0$ are all constants (though not independent). These approximations may be the simplest possible ones, but they are quite good in describing the fluctuations in the active Brownian particles, as supported later in the simulations. As the relation in Eq. \ref{sigma_eq} must be satisfied in the equilibrium limit of $v_0=0$, we have $\Delta_d^0/D_0=1$. The dependence of $D$, $\Delta_d$ and $\Delta_p$ on the norm $|{\bf p}|$ is ignored as orientation order $\langle {\bf p} \rangle=0$ throughout remains absent and the polarization fluctuation is expected to be small. Moreover, the polarization noise strength $\Delta_p$, to a good approximation, is expected to have a linear dependence on density,
\be
\Delta_p \simeq \Delta_p^0 \rho,
\ee 
where $\Delta_p^0$ is a constant. This is because the fluctuation $\sigma^2_{p_x} = \langle p_x^2 \rangle - \langle p_x \rangle^2 \sim 2 \Delta_p/D_r$ in the $x$ component of polarization density ${\bf p}$ can be written as $\sigma^2_{p_x} = \rho [\langle u_{ix}^2 \rangle - \langle u_{ix} \rangle^2] \propto \rho$ (similarly for the $y$ component) where $u_{ix}$ is the $x$-component of the orientation unit-vector of $i$th particle and therefore $\Delta_p \propto \rho$ (for details, see Appendix C). Therefore, the scaled variance as in Eq. \ref{sigma1} can be written as given below,
\begin{eqnarray}
\sigma^2(\rho) &=&  \left[ \frac{(\Delta_d^0 + \Delta_d^1 v_0^2) \rho (1-\lambda \rho)^2}{\mathcal{D}}  + \frac{\Delta_p^0 \rho v^2}{D_r^2\mathcal{D}} \right] 
\nonumber \\
&=&  \frac{(\Delta_d^0 + \Delta_d^1 v_0^2) D_r^2 \rho (1-\lambda \rho)^2 + \Delta_p^0 \rho v^2}{D_r^2 \mathcal{D}}.
\label{eq56a}
\end{eqnarray}
Now, using the previous results for the bulk velocity in the active Brownian particles, $v(\rho) \simeq v_0 (1- \lambda \rho)$  \cite{Speck_EPL2013, Speck_PRL2014} with $1/\lambda = \rho_m$ close-packing density, in Eq. \ref{sigma1}, effective diffusivity in the above equation can be written as
\begin{eqnarray}
\mathcal{D} = D_0 + \frac{v \alpha}{D_r} = D_0 \left[1+ \frac{v_0^2 (1-\lambda \rho)(1-2\lambda \rho)}{2D_r D_0}\right], \nonumber 
\end{eqnarray}
where $\alpha = ({1}/{2}) \left[ v(\rho)+\rho {dv}/{d\rho} \right] = ({v_0}/{2}) [1-2 \lambda \rho].$
Therefore, the scaled variance in Eq. \ref{eq56a} can be written as
\begin{eqnarray}
\nonumber
\sigma^2(\rho) 
&=& \rho_m \frac{(A + B P) x(1-x)^2}{1+P(1-x)(1-2x)}
\nonumber \\
&=& \rho_m \frac{(1 + A_1 v_0^2 + B P) x(1-x)^2}{1+P(1-x)(1-2x)}
\label{sigma_leading}
\end{eqnarray}
where $x =\lambda \rho = \rho/\rho_m$ is scaled density, the dimensionless parameters $A$, $A_1$ and $B$ are defined as
\bea 
A =   (1 + A_1 v_0^2);~~ A_1 = \frac{\Delta_d^1}{\Delta_d^0};~~ B = \frac{2\Delta_p^0}{D_r},
\eea
and the dimensionless scaled activity parameter
\bea
P = \frac{v_0^2}{2 D_r D_0}.
\label{P_v0}
\eea
It is customary to define another dimensionless parameter, called activity parameter or Peclet number, $Pe={v_0 \tau}/{a}$,
where the microscopic diffusive time scale $\tau = {a^2}/{D_0}$. Now, using a near-equilibrium linear-response relation $D_r = c D_0/a^2$ between the orientation (or the polarization) relaxation rate and the translational diffusion constant \cite{Stanley}, we express $P$ in terms of $Pe$,
\be 
P \simeq \frac{Pe^2}{2c},
\label{P_Pe}
\ee
where $c$ is a proportionality constant and can be estimated from simulations (see Fig. \ref{P-Pe}(a) and the corresponding discussions later).

\begin{figure}
\begin{center}
\leavevmode
\includegraphics[trim={0.001cm 0 0 0}, clip, width=8.7cm, angle=0]{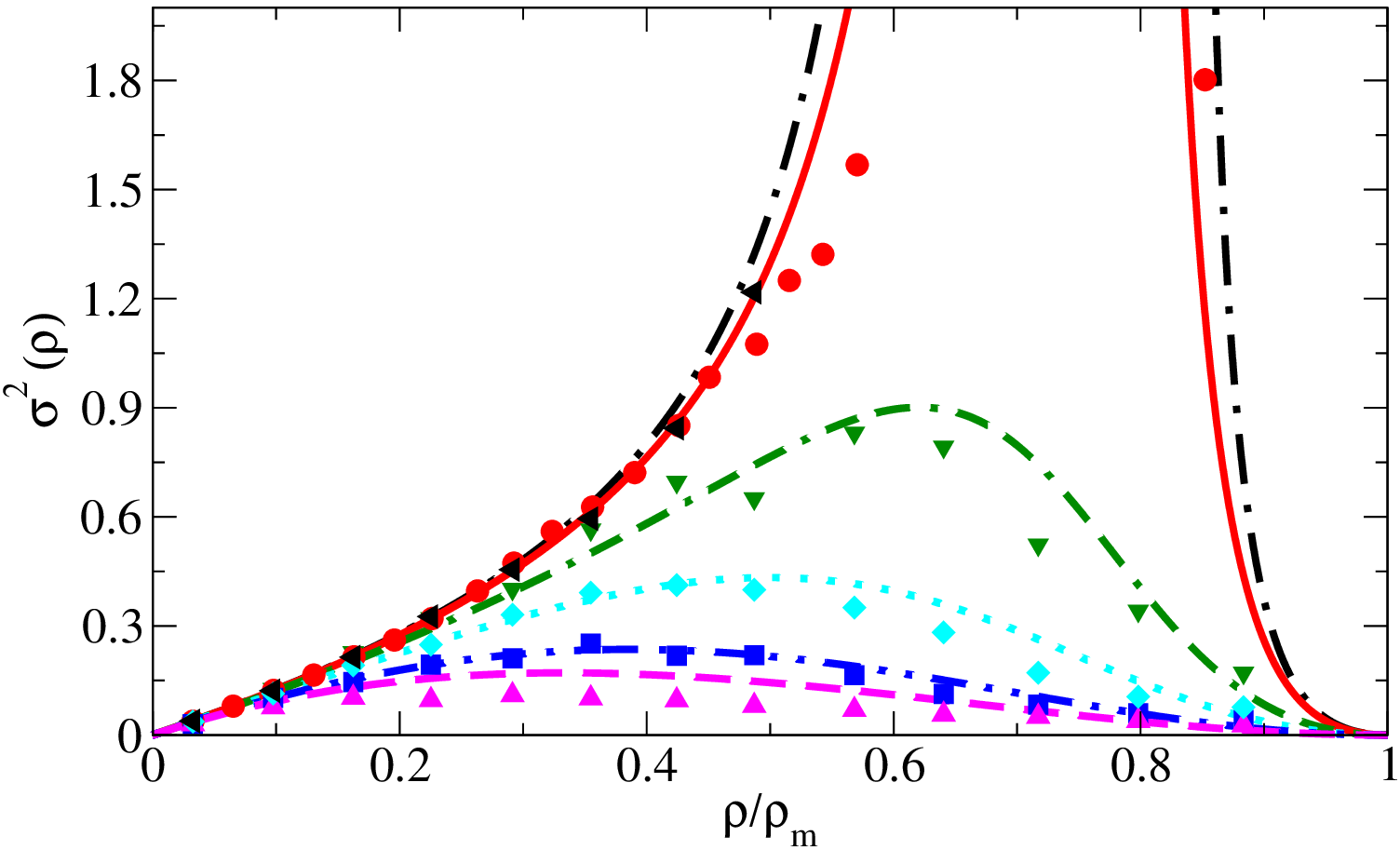}
\caption{(Color online) Simulations in the active Brownian particles. Scaled variance $\sigma^2$, for $Pe=0$ (magenta triangles), $5$ (blue squares), $10$ (sky-blue diamonds), $20$ (green inverted triangles), $50$ (red circles) and $100$ (black left-triangles), as a function of scaled density $\rho/\rho_m$, with $\rho_m \approx 1.15$, is compared with Eq. \ref{sigma3} with $P \approx 0$ (magenta dashed line), $0.5$ (blue dashed double-dotted line), $2.0$ (sky-blue dotted line), $4.3$ (green double-dashed dotted line), $8.0$ (red solid line) and $10$ (black dashed dotted line), respectively. Points - simulations, lines - theory. }
\label{fluc}
\end{center}
\end{figure}

\begin{figure}
\begin{center}
\leavevmode
\includegraphics[trim={0.001cm 0 0 0},clip,width=8.7cm,angle=0]{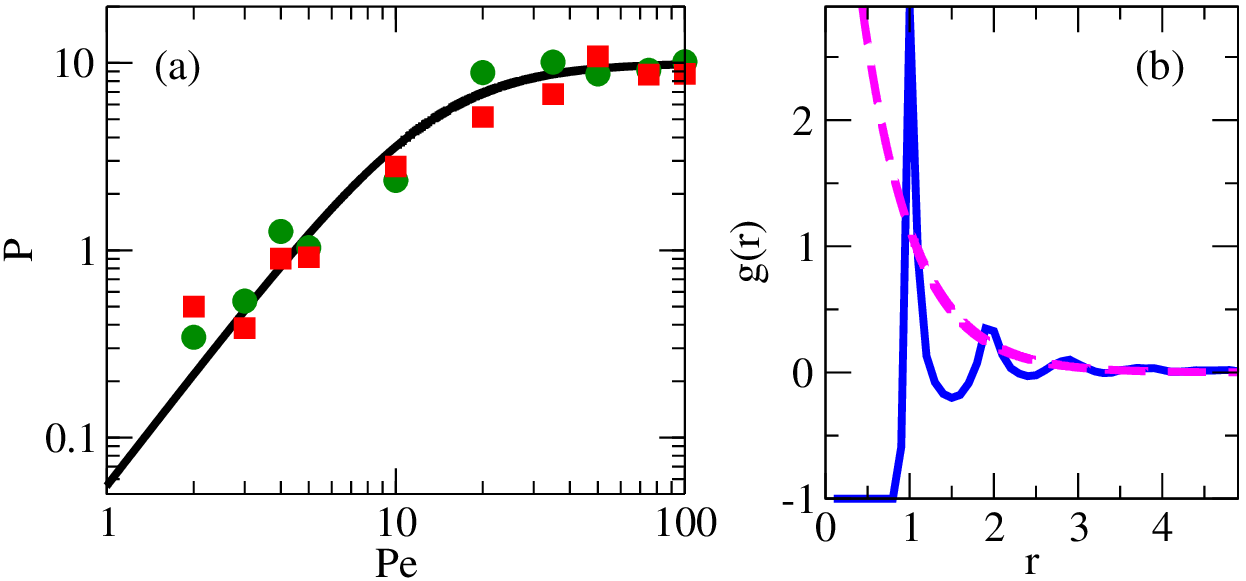}
\caption{(Color online) Panel (a): The scaled activity parameter $P$ (as defined in Eq. \ref{sigma3}) is plotted as a function of Peclet number $Pe$ for densities $\rho \approx 0.26$ (green circles) and $0.34$ (red squares). Panel (b): Pair correlation $g(r)$ (blue solid line) is plotted as a function of distance $r$ for density $\rho \approx 0.5$ and $Pe = 50$; the magenta dashed line (fitting function) shows an exponential decay of the pair correlation function at large distance, with correlation length $\xi \sim 1$ (distance is in unit of diameter $a$ of the particles). Points - simulations, lines - theory with fitting parameter.}
\label{P-Pe}
\end{center}
\end{figure}

At low activity regime $Pe \ll 1$, one can actually reduce the number of parameters in Eq. \ref{sigma_leading}, from $A_1$, $B$ and $P$ to essentially a single parameter $P$, using a constraint these parameters $A_1$, $B$ and $P$ must satisfy. It is not difficult to see that, at low density $x =\rho/\rho_m \rightarrow 0$, the particle-number distribution $P_{\cal V}({\cal N})=\exp(-\langle {\cal N} \rangle) \langle {\cal N} \rangle^{\cal N}/{\cal N}!$, for any self-propulsion $v_0$, must be Poissonian (verified in simulations; see Fig. \ref{P_n}). Therefore, the variance at low density must satisfy the constraint $\sigma^2(\rho) = \rho$, implying
\be
 A_1 v_0^2 + BP = P,
\label{constraint}
\ee
or $A_1 = {(1-B)}/{2 D_r D_0}$. Note that Eq. \ref{constraint} is exact in the leading order of self-propulsion velocity $v_0$. Using Eq. \ref{constraint} in Eq. \ref{sigma_leading}, we finally obtain the scaled variance as a function of scaled density $x=\rho/\rho_m$, 
\be
\sigma^2(\rho) = \rho_m \frac{(1 + P) x(1-x)^2}{1+P(1-x)(1-2x)},
\label{sigma3}
\ee 
which essentially represents a one-parameter family of curves (see $\sigma^2(\rho)$ as a function of $\rho$ for various $P$ in Fig. \ref{fluc}) with the scaled activity parameter $P \simeq Pe^2/2c$ as in Eq. \ref{P_Pe}. 
Interestingly, as we find below in the simulations of the active Brownian particles, the form of the variance in Eq. \ref{sigma3} indeed captures quite well the broad features of particle-number fluctuations even when activity is moderately large $Pe \gg 1$.

Now we show, using the form of the scaled variance in Eq. \ref{sigma3}, how the scaled activity parameter $P$ can be estimated from the simulations of the active Brownian particles. This is done essentially by fitting Eq. \ref{sigma3} for a suitable choice of the fitting parameter $P$. In Fig. \ref{fluc}, we plot scaled variance $\sigma^2$ as a function of $x=\rho/\rho_m$, obtained from simulations for various $Pe=0$ (magenta triangles), $5$ (blue squares), $10$ (sky-blue diamonds), $20$ (green inverted triangles), $50$ (red circles) and $100$ (black left-triangles), and then fit the curves with Eq. \ref{sigma3} by suitably choosing $P \approx 0$ (magenta dashed line), $0.5$ (blue dashed double-dotted line), $2.0$ (sky-blue dotted line), $4.3$ (green double-dashed dotted line), $8.0$ (red solid line) and $10$ (black dashed dotted line), respectively. To find the dependence of $P$ on $Pe$, we numerically calculate $P$ as a function of $Pe$, by solving for $P$ where we use a particular value of scaled density $x$ and variance $\sigma^2$ in Eq. \ref{sigma3}. In Fig. \ref{P-Pe}(a), we plot $P$ as a function of $Pe$, for a set of two densities $\rho=0.26$ (green circles) and $0.34$ (red squares). The function fits quite well with the form $P \simeq Pe^2/(2c+\kappa Pe^2)$ [see the black solid line in Fig. \ref{P-Pe}(a)] where $c \approx 9$, implying a somewhat larger coarse-grained relaxation rate $D_r$ for the polarization field than that estimated previously \cite{Baskaran_PRL2013}, and $\kappa \approx 0.1$. In other words, at smaller activity regime $Pe \lesssim 20$, the scaled activity parameter $P \simeq (Pe)^2/2c$ varies quadratically with $Pe$ as in Eq. \ref{P_Pe}. However, for very large activity $Pe \gsim 20$, as discussed above, the scaled activity parameter $P \simeq 1/\kappa$ eventually saturates.

\section{Density fluctuations and Nonequilibrium thermodynamics}
\label{noneq_therm}

\subsection{Chemical potential and free energy function}

In this section, we calculate, using the analytic form of the variance in Eq. \ref{sigma3}, nonequilibrium chemical potential $\mu(\rho)$ and free energy density function $f(\rho)$. We use the fluctuation-response relation Eq. \ref{FR1}, change the density variable $\rho$ to a scaled density $x=\rho/\rho_m$ and integrate w.r.t. the scaled density $x$,
\be 
\frac{d\rho}{d\mu} = \sigma^2(\rho) \Rightarrow \frac{d x}{d\mu} = \frac{(1 + P) x(1-x)^2}{1+P(1-x)(1-2x)},
\ee
to obtain nonequilibrium chemical potential as a function of the scaled density $x$,
\bea
\mu(x) = \frac{1}{1 + P} \left[ (P-1) \ln (1-x)+(P+1) \ln x
+ \frac{1}{1-x} \right] \nonumber
\eea
which, upon substituting $x=\rho/\rho_m$, gives chemical potential as a function of density $\rho$
\bea
\mu(\rho) = \frac{1}{1 + P} [ (P-1) \ln (1-\frac{\rho}{\rho_m})+(P+1) \ln (\frac{\rho}{\rho_m}) 
\nonumber \\ 
+ \frac{1}{1-\frac{\rho}{\rho_m} } ].~ 
\label{mu-SM}
\eea
Now, integrating chemical potential $\mu(\rho)$ w.r.t. density $\rho$, we get free energy density function
\bea
f(\rho) = \int \mu d\rho = \rho_m \int \mu(x) dx ~~~~~~~~~~~~~~~~~~~~~~~~~~~~~~~~~~~~
\nonumber \\
= \rho_m \frac{[P (x-1)-x] \ln (1-x) + x [(P+1) \ln x -2 P]}{1 + P},~~~
\label{f-SM}
\eea
which has the dimension of density and, upon substituting $x=\rho/\rho_m$, gives free energy density as a function of density $\rho$.

\subsection{Subsystem particle-number distributions}

Nonequilibrium free energy density function in Eq. \ref{f-SM}, being a large deviation function, and nonequilibrium chemical potential in Eq. \ref{mu-SM}, together, govern the particle-number fluctuation in the system. Therefore, based on the analytical result of subsystem particle-number distribution in Eq. \ref{P_N_additivity} which can be explicitly calculated now using Eqs. \ref{sigma3}, \ref{mu-SM} and \ref{f-SM} (see section \ref{sec-additivity}), we finally test in this section the predictions of additivity concerning density fluctuations in actual simulations of the active Brownian particles. The simulations are performed in the fluid phase, which is much away from criticality, where Eq. \ref{sigma3} is expected to hold.

\begin{figure}
\begin{center}
\leavevmode
\includegraphics[trim={0.001cm 0 0 0},clip,width=8.5cm,angle=0]{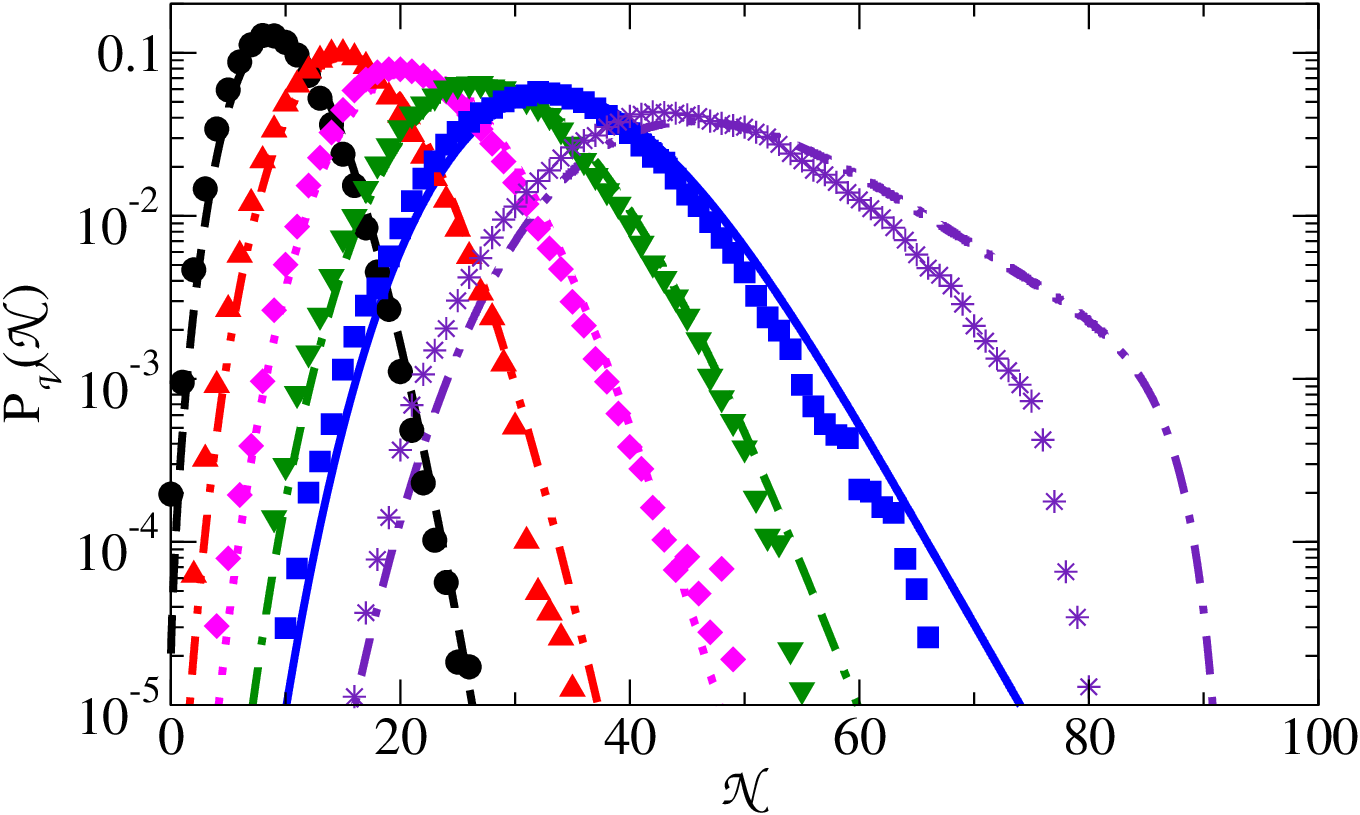}
\caption{(Color online) Subsystem particle-number distributions for activity parameter $Pe = 50$, obtained from simulations (points), are compared with theory Eq. \ref{P_N_additivity} (lines) with corresponding scaled activity parameter $P = 8$, for densities $\rho \approx 0.11$ (black circles, black dashed line), $0.19$ (red triangles, red dashed double-dotted line), $0.26$ (magenta diamonds, magenta dotted line), $0.34$ (green inverted triangles, green doubled-dashed dotted line), $0.41$ (blue squares, blue solid line) and $0.56$ (violet asterisks, violet dashed dotted line). }
\label{P_n}
\end{center}
\end{figure}

\begin{figure}
\begin{center}
\leavevmode
\includegraphics[trim={0.001cm 0 0 0},clip,width=8.5cm,angle=0]{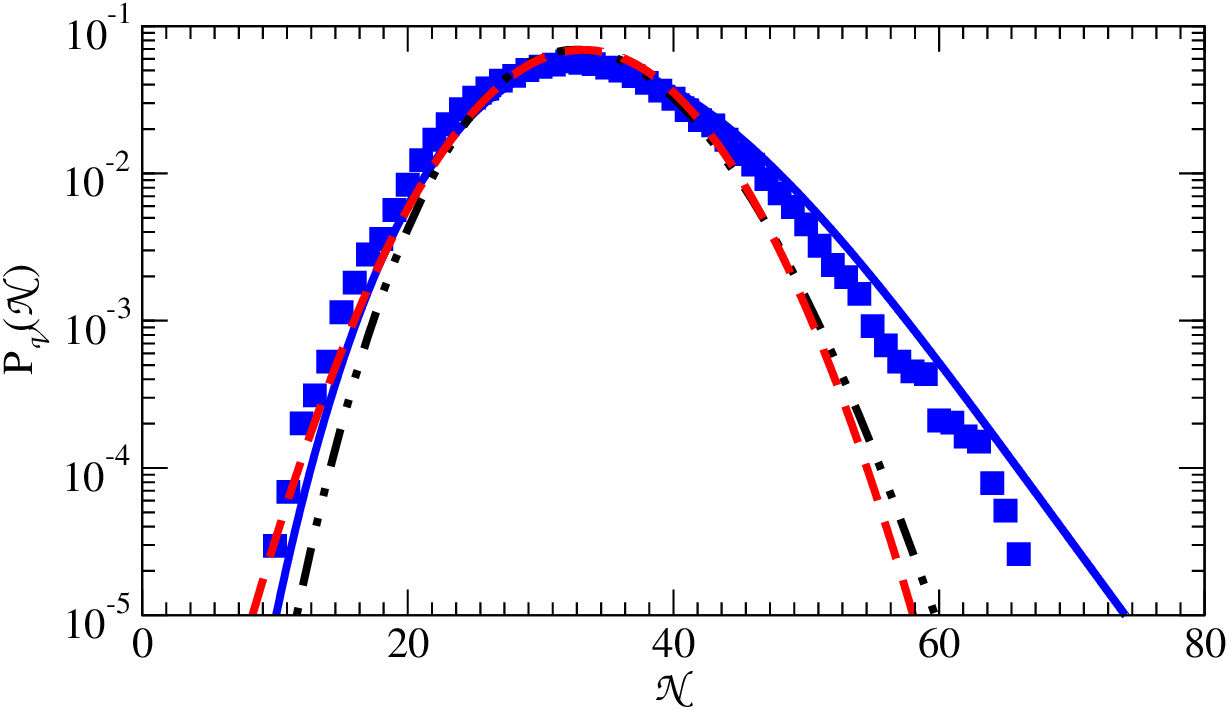}
\caption{(Color online) At moderately large density $\rho \approx 0.41$ and large activity $Pe=50$, subsystem particle-number distribution $P_{\cal V}({\cal N})$ (blue squares), which deviates from Poisson (black dashed double-dotted line) as well as Gaussian (red dashed line) distributions, is quite well captured by theory Eq. \ref{P_N_additivity} (blue solid line).  }
\label{fluc_poisson_gaussian}
\end{center}
\end{figure}

In simulations, we calculate subsystem particle-number distributions $P_{\cal V}({\cal N})$ in a subsystem (${\cal V}=9 \times 9$ in units of $a$) where the rest of the system ($V=100 \times 100$) acts as a particle {\it reservoir} of chemical potential $\mu(\rho)$. In Fig. \ref{P-Pe}(b), we plot pair-correlation function $g({\bf r})= \sum_{i \ne 1} \langle \delta ({\bf r} - {\bf R}_i(t)) \rangle$ as a function of distance $r$ at a moderately high density $\rho \approx 0.5$ and $Pe = 50$ where correlation length $\xi \sim {\cal O}(a)$, much smaller than the subsystem size. In Fig. \ref{P_n}, subsystem number-distributions $P_{\cal V}({\cal N})$ obtained from simulations (points) at $Pe=50$ are compared with theory Eq. \ref{P_N_additivity} (lines) at the corresponding scaled activity $P=8$, for several densities  $\rho \approx 0.11$ (black circles, black dashed line), $0.19$ (red triangles, red dashed double-dotted line), $0.26$ (magenta diamonds, magenta dotted line), $0.34$ (green inverted triangles, green doubled-dashed dotted line), $0.41$ (blue squares, blue solid line) and $0.56$ (violet asterisks, violet dashed dotted line). Agreement between simulations and theory, even at quite large density $\rho \approx 0.41$, is reasonably good. Note that, provided the variance $\sigma^2(\rho)$  as a function of density $\rho$ (as in Eq. \ref{sigma3}), there is no fitting parameter in the distribution functions $P_{\cal V}({\cal N})$ plotted in Figs. \ref{P_n} and \ref{fluc_poisson_gaussian}. Expectedly, the distributions are Poissonian at low densities. However, the distributions become increasingly non-Poissonian, or non-Gaussian, with increasing density and activity. To emphasize this point, in Fig. \ref{fluc_poisson_gaussian}, we show that, for moderately large density $\rho \approx 0.41$ and large activity $Pe=50$, the particle-number distribution function $P_{\cal V}({\cal N})$ in simulations (blue squares) indeed deviates from the corresponding Poisson (black dashed double-dotted line) as well as Gaussian (red dashed line) distributions. Even then, the numerically obtained distribution (blue diamonds) is indeed quite well described by the analytically obtained distribution Eq. \ref{P_N_additivity} (blue solid line), thus validating additivity, at least in the homogeneous fluid phase which is sufficiently away from criticality.

However, upon approaching closer to the criticality, some discrepancies arise between analytic theory and simulations, presumably due to the linear analysis of Eqs. \ref{Hydro1} and \ref{Hydro2} and finite-size effects. That the linear analysis breaks down at density $\rho \approx 0.56$ is evident from Fig. \ref{fluc} where simulation results (for $Pe = 50$) start deviating from the analytic expressions of Eq. \ref{sigma3} (for corresponding $P=8$). On the other hand, the finite-size effects originate from the facts that the boundary correlations between subsystem and system (due to increasing correlation length) increase while approaching criticality and the ratio between system and subsystem as well as their individual sizes are finite.

\subsection{Phase transition}

\begin{figure}
\begin{center}
\leavevmode
\includegraphics[trim={0.001cm 0 0 0},clip,width=8.3cm,angle=0]{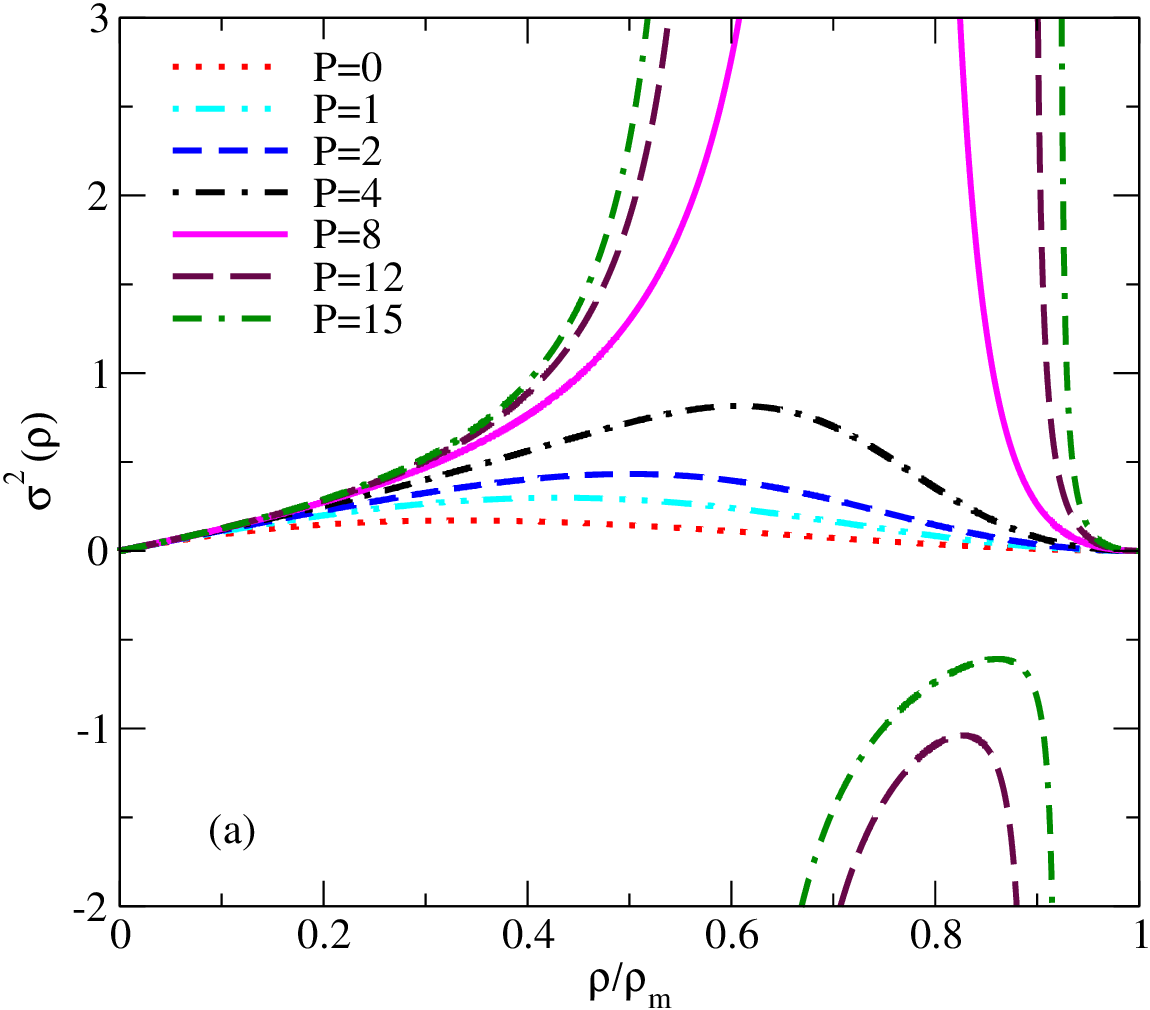}
\includegraphics[trim={0.001cm 0 0 0},clip,width=8.3cm,angle=0]{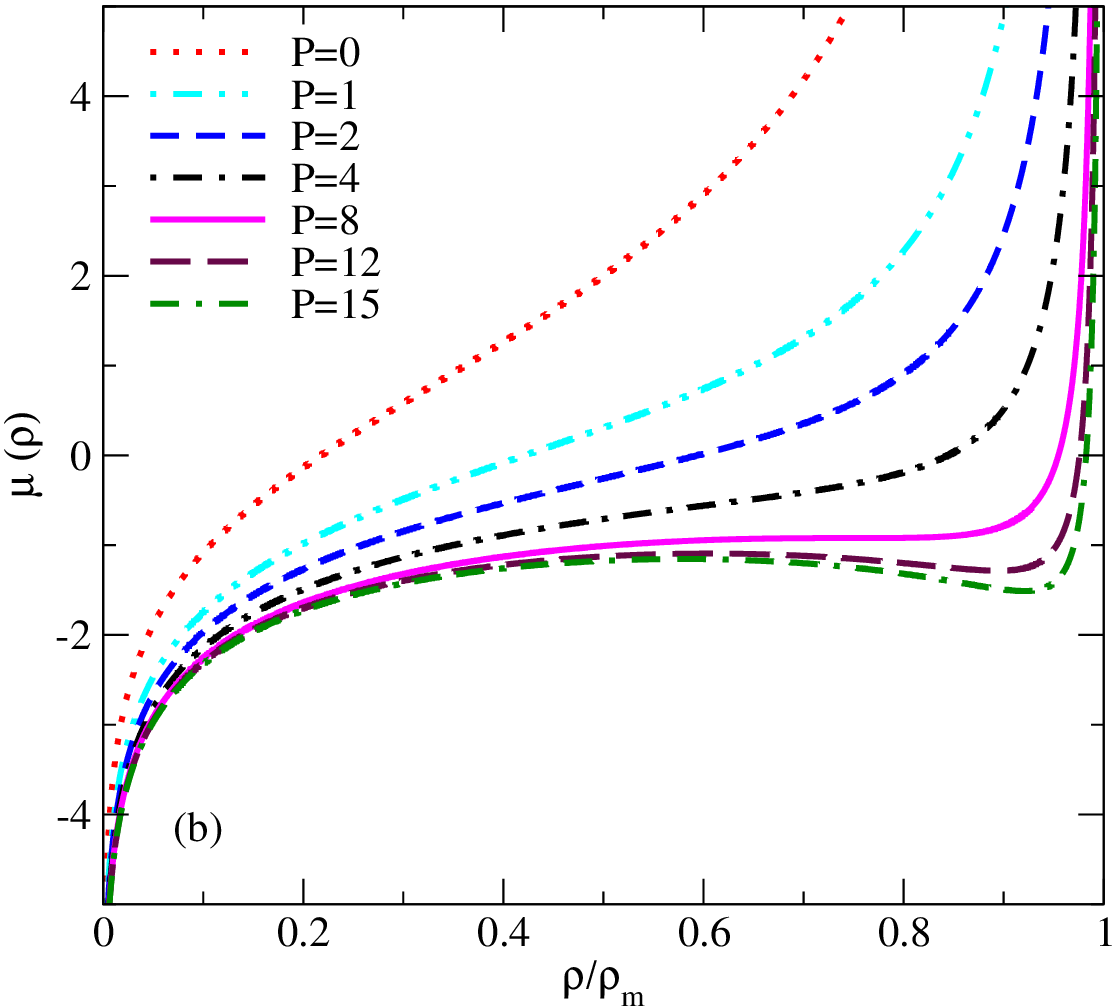}
\caption{Scaled variance $\sigma^2(\rho)$ (Eq. \ref{sigma3}) and corresponding chemical potential $\mu(\rho) = \int^{\rho} 1/\sigma^2 d \rho$ (Eq. \ref{mu-SM}) are plotted in panels (a) and (b), respectively as a function of the scaled density $\rho/\rho_m$ for various values of the scaled activity parameter $P = 0$, $1$, $2$, $4$, $8$, $12$ and $15$. Chemical potential becomes a nonmonotonic function of density beyond a critical value of the scaled activity parameter, $P>P_c=8$.}
\label{fig10}
\end{center}
\end{figure}

Based on the analysis in the previous sections, one can now formulate a theory of phase transition in the active Brownian particles. We note that the functional form of the scaled variance as in Eq. \ref{sigma3} has many interesting implications. In the regime of large activity where Peclet number $Pe \gg 1$ (or $P \gg 1$), the scaled variance $\sigma^2(\rho)$ is independent of $P$, which  is in quite good agreement with simulations (see Fig. \ref{fluc} for $Pe=50$ and $100$) where $\sigma^2$ for any $\rho$ almost saturates at large Peclet number. Moreover, the denominator in Eq. \ref{sigma3} has two roots 
$$
x_{2,1} = \frac{3}{4} \pm \frac{\sqrt{P^2-8P}}{4P}.
$$ 
In Fig. \ref{fig10}(a), we plot the scaled variance (as in Eq. \ref{sigma3}) and,  in Fig. \ref{fig10}(b), chemical potential (as in Eq. \ref{mu-SM}) as a function of scaled density $x=\rho/\rho_m$, with $\rho_m \approx 1.15$, for various values of scaled activity $P=0$, $1$, $2$, $4$, $8$, $12$ and $15$. Below a critical value of the scaled activity $P < P_c=8$ (corresponding to activity $Pe = 50$ in actual simulations), the variance remains positive in the full interval $0 \le x \le 1$. 
On the other hand, above a critical value of scaled activity $P > P_c$, the variance becomes negative in the interval $x_1 < x < x_2$ and consequently chemical potential becomes nonmonotonic function of density, which is {\it not} physical and implies onset of phase coexistence. The coexisting densities can, in principle, be calculated using a Maxwell construction on chemical potential $\mu(\rho) = \int^{\rho} 1/\sigma^2 d\rho$ (Eq. \ref{mu-SM}) or on free energy density function $f(\rho) = \int^{\rho} \mu d\rho$ (Eq. \ref{f-SM}). Presently, the Maxwell construction is however not expected to give an accurate estimate of the coexisting densities as our theory (Eq. \ref{sigma3} and consequent expressions in Eqs. \ref{mu-SM} and \ref{f-SM}) have been derived using a linearized hydrodynamics and a near-equilibrium analysis, which do not capture well the fluctuations in the high activity regime.

Somewhat surprisingly, our theory however predicts quite accurately the critical density $\rho_c$ where compressibility $d\rho/d\mu$ diverges; $\rho_c \approx 0.86$, or critical packing fraction $\phi_c \approx 0.7$, obtained from theory is in excellent agreement with simulations \cite{Baskaran_PRL2013}. Moreover, we find that compressibility diverges as $d\rho/d\mu = \sigma^2 \sim 1/(\rho-\rho_c)^{\delta-1}$, or equivalently chemical potential vanishes as $\mu \sim (\rho-\rho_c)^{\delta}$, with $\delta=3$; correlation length diverges as $\xi \sim (\rho-\rho_c)^{-\nu_h}$ with mean-field $\nu_h=1$. Clearly, on the mean-field level, the exponents are in accordance with Ising universality; $P$ and $\mu$ are analogous to temperature and magnetic field, respectively.

\section{Summary and concluding perspective}  
\label{Summary}

In this paper, using a nonequilibrium fluctuation-response relation - a direct consequence of an additivity property, we formulate a thermodynamic theory for self-propelled particles in the context of a particular model system consisting of active Brownian particles. From the fluctuation-response relation, we demonstrate that subsystem particle-number distributions, which, being related to the density large deviation function and thus analogous to equilibrium free energy, can help us to characterize macroscopic properties in self-propelled particles, in a unified statistical mechanics framework, in terms of a nonequilibrium chemical potential. Analogous to phase transition in equilibrium, as density and activity (Peclet number) increase, chemical potential becomes nonmonotonic function of density, indicating onset of a gas-liquid phase coexistence.

Importantly, the formalism developed here is solely based on characterization of the variance of subsystem particle number, which is directly related to the two-point (equal-time) density correlations or the structure factor. Provided that one calculates the structure factor accurately, our theory can lead to verifiable predictions concerning the density fluctuations. However, analytically calculating structure factor in a many-particle system is not an easy task. To this end, in the first step, we have calculated, though within a linearized fluctuating hydrodynamics, the structure factor in a microscopic model system of active Brownian particles and, consequently, the variance of subsystem particle-number as a function of density. Our studies of fluctuating hydrodynamics provide some insights in characterizing the noises in the hydrodynamic equations, done in the context of active Brownian particles. For this purpose, we have used a near-equilibrium analysis, which, though approximate, captures reasonably well the broad features of the particle-number fluctuations even in the far-from-equilibrium regime where activity is moderately large.

Furthermore, in the second step, from the functional dependence of the variance of subsystem particle-number on density and then using additivity, we have calculated the subsystem particle-number distribution function and have compared the analytically obtained distribution functions with that obtained from simulations in the active Brownian particles. The agreement between theory and simulations is quite good, except some deviations at the tails. The deviations increase while approaching criticality, indicating the following reasons for these deviations. Firstly, the linear analysis used here breaks down in the regime of high densities and the high activities, where nonlinear effects can induce nontrivial fluctuations; consequently, chemical potential and free energy function obtained from the linear analysis cannot capture the density fluctuations well. Secondly, there can be significant finite-size effects, originating from the increasing boundary correlations between subsystem and system upon approaching criticality and due to the finite ratio between system and subsystem (simulations here are performed upto total particle number $N > 10^4$ and roughly for ratio $1:123$ between subsystem and system volumes). Thus, larger scale simulations, though computationally difficult at this stage, would be quite useful for more accurate verification of the predictions of additivity.

For simplicity, here we have restricted ourselves to a particular model system of active Brownian particles. However, the thermodynamic formalism developed here is quite general and could be extended to other active-matter systems, e.g., models with run-and-tumble Bacterial dynamics or Vicsek model \cite{Viscek_PRL1995} and its variants. Moreover, even in the active Brownian particles, it would be quite interesting, though challenging, to go beyond the linear hydrodynamic regime by allowing nonlinear gradient terms (as in \cite{Cates_NatCom2014}) or self-advective terms (as in \cite{Viscek_PRL1995, Bar}), which may be relevant in the large activity regime where fluctuations are large.

From the overall perspective, we believe additivity could be the missing link, providing a unified characterization of a broad range of phenomena in the self-propelled particles observed in the past. Also, it will be interesting to explore if additivity holds in the phase with ``giant number'' fluctuations which many self-propelled particle systems exhibit or in the presence of inhomogeneities, e.g., a confining potential as in a box with hard walls \cite{Tailleur_Nature2015}, etc.

\section{Acknowledgement}

We thank Pradeep K. Mohanty for useful discussions. SC and PP acknowledge Thematic Unit of Excellence on Computational Materials Science, funded by the Department of Science and Technology (India) for computational facility used in the present study. SM acknowledges financial support (under INSPIRE faculty award 2012) from the Department of Science and Technology (India).

\section*{APPENDIX A: Additivity and fluctuation-response relation}

Here we show how additivity, as in \ref{additivity1} in the main text, directly leads to the fluctuation-response relation as in Eq. \ref{FR1} in the main text. Provided additivity property is satisifed, the subsystem particle-number distribution can be written , in the thermodynamic limit, as
\be 
P_{\cal V}({\cal N}) = \frac{1}{\cal Z} W_{\cal V}({\cal N}) e^{\mu {\cal N}},
\ee 
where ${\cal W}$ is the weight factor for the respective subsystem, $\mu$ is a chemical potential and the normalization constant ${\cal Z}$ is given by
\be 
{\cal Z}(\mu) = \sum_{{\cal N}=0}^{\infty} W_{\cal V}({\cal N}) e^{\mu {\cal N}}.
\label{normalization}
\ee 
Now, the average particle number can be calculated by taking a derivative of logarithm of the normalization constant w.r.t. $\mu$,
\be 
\langle {\cal N} \rangle = \frac{d (\ln {\cal Z})}{d \mu}.
\label{avg_N}
\ee
By taking another derivative of Eq. \ref{avg_N} w.r.t. $\mu$, one can immediately relate compressibility to the fluctuation,
\be 
\frac{d {\langle \cal N \rangle}}{d \mu} = \frac{d^2 (\ln {\cal Z})}{d \mu^2} = \langle {\cal N}^2 \rangle - \langle {\cal N} \rangle^2,
\ee
where, in the last step, we have used Eq. \ref{normalization}. Dividing both side of the above equation by the subsystem volume ${\cal V}$, we get, in the limit of large ${\cal V}$, the fluctuation-response relation as in Eq. 3 in the main text,
\be 
\frac{d \rho}{d \mu} = \sigma^2,
\ee
where the scaled variance is defined as 
$$
\sigma^2 = \lim_{{\cal V} \rightarrow \infty} \frac{(\langle {\cal N}^2 \rangle - \langle {\cal N} \rangle^2)}{{\cal V}}.
$$

\section*{APPENDIX B: Calculation of Structure factor in the active Brownian particles within linearized fluctuating hydrodynamics}

We consider the following fluctuating hydrodynamic equations, as considered in the main text, for self-propelled particles (SPP)
\bea
\label{SHydro1}
\partial_t \rho &=& - {\bf \nabla}.\left[ v(\rho) {\bf p} - D(\rho) {\bf \nabla }\rho+ {\bf f}_d \right],
\\
\label{SHydro2}
\partial_t{\bf{p}} &=& -D_r{\bf{p}}-\frac{1}{2}{\bf{\nabla}}\left(\rho v \right)+ K\nabla^2{\bf{p}}+ {\bf f}_p,
\eea
and perform linear analysis along the lines of Ref. \cite{Fily_Marchetti_PRL2012}. We transform the variable $\theta({\bf r})=\nabla.{\bf p}$, expand the nonlinear terms upto linear order of $\delta \rho$ and $\delta {\bf p}$, where $\delta \rho = \rho - \rho_0$, $\delta {\bf p} = {\bf p} - {\bf p}_0$, $\delta \theta = \nabla. (\delta {\bf p})$ with ${\rho}_0$ and ${\bf p}_0=0$ average density and polarization profile, to obtain
\begin{eqnarray}
\label{eq3}
\partial_t \delta \rho({\bf{r}},t) = -v(\rho_0) \delta \theta({\bf{r}},t) + D(\rho_0) \nabla^2 \delta \rho({\bf{r}},t) - {\bf{\nabla}}.{\bf f}_d,
\\
\label{eq4}
\partial_t \delta \theta({\bf{r}},t)=-D_r\delta \theta({\bf{r}},t)-\alpha(\rho_0) \nabla^2\delta \rho({\bf{r}},t)~~~~~~~~~~~~~~
\nonumber \\
+ K\nabla^2\delta \theta({\bf{r}},t) + {\bf{\nabla}}.{\bf f}_p,
\end{eqnarray}
where $\nabla.(v {\bf p}) \simeq v \delta \theta$, $\nabla (\rho v) \simeq 2 \alpha(\rho_0) \nabla (\delta \rho)$ with 
$$
2 \alpha(\rho_0) = \frac{d}{d\rho_0} [\rho_0 v(\rho_0)] = v(\rho_0) + \rho_0 \frac{dv(\rho_0)}{d\rho_0}.
$$
Using Fourier amplitudes
\begin{eqnarray}
\label{eq5}
\delta\tilde{\rho}({\bf{q}},\omega)&=&\int_{\bf{r}}\int_t e^{-i{\bf{q}}.{\bf{r}}} e^{-i\omega t}~\delta \rho({\bf{r}},t) d{\bf r} dt,
\\
\label{eq6}
\delta\tilde{\theta}({\bf{q}},\omega)&=&\int_{\bf{r}}\int_t e^{-i{\bf{q}}.{\bf{r}}} e^{-i\omega t}~\delta \theta({\bf{r}},t) d{\bf r} dt,
\end{eqnarray}
and reverting back to global density $\rho_0 = \rho$ (for notational simplicity), Eqs.(\ref{eq3}) and (\ref{eq4}) can be written as
\begin{eqnarray}
\label{eq7}
[i\omega+q^2D(\rho)]\delta\tilde{\rho} + v(\rho) \delta \tilde{\theta}&=& -i{\bf{q}}.\tilde{{\bf f}}_d \\
\label{eq8}
\alpha(\rho) q^2\delta\tilde{\rho} - (D_r + K q^2 + i\omega)\delta \tilde{\theta}&=& -i{\bf{q}}.\tilde{\bf f}_p.
\end{eqnarray}
Solving for the Fourier modes, we get
\begin{eqnarray}
\nonumber
\left[ {\begin{array}{cc}
\delta\tilde{\rho}\\
\delta\tilde{\theta}\\
\end{array} } \right]= -i{\bf q}.
\left[ {\begin{array}{cc}
q^2D(\rho)+ i\omega & v(\rho) \\
\alpha q^2 & -( D_r + Kq^2 + i\omega)\\
\end{array} } \right]^{-1}
\left[ {\begin{array}{cc}
\tilde{{\bf f}}_d\\
\tilde{{\bf f}}_p\\
\end{array} } \right]
\end{eqnarray}
and therefore
\begin{eqnarray}
\label{eqn11}
\delta\tilde{\rho}({\bf{q}},\omega) = \frac{i}{{\rm det}(M)}[(D_r + Kq^2 + i\omega){\bf q}.\tilde{{\bf f}}_d + v {\bf q}.\tilde{{\bf{f}}}_p]
\end{eqnarray}
with
\begin{equation}
M = \left[ {\begin{array}{cc}
 i \omega+q^2D(\rho) & v(\rho)\\
\alpha q^2 & -(Kq^2+D_r + i\omega)\\
\end{array} } \right].
\end{equation}
Using the noise correlations, $\langle |{\bf q}.\tilde{{\bf f}}_d|^2\rangle = 2 V \Delta_d q^2$, $\langle |{\bf q}.\tilde{{\bf f}}_p |^2\rangle = 2V \Delta_p q^2$ and $\langle({\bf q}.\tilde{{\bf f}}_d^*)({\bf q}.\tilde{{\bf f}}_p)\rangle = \langle({\bf q}.\tilde{{\bf f}}_d)({\bf q}.\tilde{{\bf f}}_p^*)\rangle = 0$, we obtain dynamic structure factor 
\begin{eqnarray}
\nonumber
S({\bf{q}},\omega) &=& \langle |\delta \tilde{\rho}({\bf q},\omega)|^2 \rangle
\nonumber \\
&=&\frac{\lbrace\omega^2+(D_r+Kq^2)^2\rbrace\langle |{\bf q}.\tilde{{\bf f}}_d|^2\rangle+v^2\langle |{\bf q}.\tilde{{\bf f}}_p|^2\rangle}{\vert {\rm det}(M)\vert^2} 
\nonumber \\
&=& \frac{2 q^2 V}{\vert {\rm det}(M)\vert^2}[\Delta_d \lbrace \omega^2 + (D_r + Kq^2)^2 \rbrace + \Delta_p v^2]~~~~~~
\end{eqnarray}
where
\begin{eqnarray}
\label{eq14}
\vert {\rm det}(M)\vert^2=(\omega^2-a)^2+\omega^2b^2
\end{eqnarray}
with $$a=q^2[D_r {\cal D}(\rho) + D(\rho) Kq^2],$$ $$b=D_r + q^2[K+D(\rho)],$$ and $${\mathcal{D}}(\rho)=D(\rho) + v(\rho) \alpha(\rho)/D_r.$$ The Static Structure factor  can be computed as $S({\bf{q}})=({1}/{2\pi}) \int_{-\infty}^\infty S({\bf{q}},\omega)~d\omega$. Now using the following equalities,
\begin{eqnarray}
\nonumber
\int_{-\infty}^\infty \frac{d\omega}{(\omega^2-a)^2+\omega^2b^2}=\frac{\pi}{ab},
\\
\int_{-\infty}^\infty \frac{\omega^2 d\omega}{(\omega^2-a)^2+\omega^2b^2}=\frac{\pi}{b},
\end{eqnarray}
we obtain $S({\bf{q}})=S_1({\bf{q}})+S_2({\bf{q}})$ where
\begin{eqnarray}
\nonumber 
S_1({\bf{q}})= V \frac{2 q^2 \Delta_d}{2\pi}\left[ \frac{\pi}{b}+(D_r+Kq^2)^2 \frac{\pi}{ab}\right]\\
=  \frac{V\Delta_d q^2}{D_r+q^2(K+D)} ~~~~~~~~~~~~~~~~~~~
\nonumber \\
+ \frac{V\Delta_d q^2 (D_r+Kq^2)^2}{q^2[D_r{\cal D} + D Kq^2][D_r+q^2(K+D)]}  
\end{eqnarray}
and
\begin{eqnarray}
\nonumber 
S_2({\bf{q}})&=& V \frac{2q^2 \Delta_p v^2}{2\pi} \frac{\pi}{ab}\\
\label{eq21}
&=& V \frac{\Delta_p v^2}{[D_r{\cal D} + D Kq^2][D_r+q^2(K+D)]}.
\end{eqnarray}
The structure factor $S({\bf q}=0)=S_1(0) + S_2(0)$ is related to variance $\sigma_{\cal V}^2(\rho)=\langle {\cal N}^2 \rangle - \langle {\cal N} \rangle^2$ of number of particles ${\cal N} = \int_{\cal V} \rho(\bf r) d{\bf r}$ in a {\it subvolume} ${\cal V}$ which can be written as
\begin{eqnarray}
\label{eq55}
\sigma_{\cal V}^2(\rho) = S({\bf q}=0) = {\cal V} \left[ \frac{\Delta_d}{\mathcal{D}(\rho)}  + \frac{\Delta_p v^2(\rho)}{D_r^2 \mathcal{D}(\rho)} \right], \end{eqnarray}
the desired expression in the main text.

\subsection*{APPENDIX C: Polarization fluctuations in the active Brownian particles}

As defined in the equations of motion for the active Brownian particles in the main text, ${\bf u}_i \equiv \{u_{ix, iy}\} = \{\cos \phi_i, \sin \phi_i\}$ the orientation unit vector for the $i$th Brownian particle. From the definition of the polarization density ${\bf p}({\bf r},t) = \sum_{i} \delta ({\bf r}-{\bf R}_i(t)) {\bf u}_i(t)$, we can express total polarization ${\bf P}_{\Delta V}$, in a small volume $\Delta V$, as
\be 
{\bf P}_{\Delta V} = \sum_{i \in \Delta V} {\bf u}_i(t)
\ee
where the sum is over $\Delta N$ number of particles in the volume $\Delta V$ so that 
\be 
{\bf p} = \lim_{\Delta V \rightarrow 0} \frac{{\bf P}_{\Delta V}}{\Delta V}.
\ee 
Now, using the central limit theorem (CLT), one can estimate the fluctuation or the variance of ${\bf P}_{\Delta V} \equiv \{ P^x_{\Delta V}, P^y_{\Delta V}\}$, which is the sum of $\Delta N$ random variables (i.e., the sum of random orientation unit vectors of $\Delta N$ particles in volume $\Delta V$) where the variance of the $x$ and $y$ components of each orientation unit vector ${\bf u_i}$, for any $i$, are calculated to be constant. Consequently, the variance of the $x$ and $y$ components of polarization density ${\bf p}$ can be calculated as given below,
\bea
\sigma^2_{p_{ix}} &=& \langle p_{ix}^2 \rangle - \langle p_{ix} \rangle^2  
\nonumber \\
&=& \lim_{\Delta V \rightarrow 0} \frac{\langle (P^x_{\Delta V})^2 \rangle - \langle (P^x_{\Delta V}) \rangle^2}{\Delta V}
\nonumber \\
&=& \lim_{\Delta V \rightarrow 0} \frac{\langle \Delta N \rangle}{\Delta V} [\langle u_{ix}^2 \rangle - \langle u_{ix} \rangle^2] \propto \rho,
\eea
and similarly 
\be 
\sigma^2_{p_{iy}} \propto \rho.
\ee

\subsection*{APPENDIX D: Structure factor in the limit of quasistatic polarization field ($\Delta_p=0$)}

In the quasi-static limit of polarization field, by setting $\partial_t {\bf p}=0$, $K=0$ and polarization noise strength $\Delta_p=0$ in Eq. \ref{SHydro2} and then substituting ${\bf p}$ in Eq. \ref{SHydro2} \cite{Cates_EPL2013, Cates_condmat2015}, we get an effective evolution equation for density field,
\bea
\partial_t \rho &=& - {\bf \nabla}.\left[- v \frac{\nabla(v \rho)}{2 D_r} - D {\bf \nabla} \rho + {\bf f}_d \right] 
\\
&=& -\nabla . [\tilde v \rho - \tilde D \nabla \rho + {\bf f}_d]
\eea 
where effective velocity $\tilde v (\rho)=- v\nabla v/2D_r$ and effective diffusivity $\tilde D = D+v^2/2D_r$. Now we perform linear analysis  of fluctuation $\delta \rho = \rho - \rho_0$ around the average density $\rho_0$,
\bea
\partial_t \delta \rho = -\nabla .[-{\cal D}(\rho_0) \nabla \delta \rho + {\bf f}_d] 
\eea 
where effective diffusivity 
\be 
{\cal D} = \tilde D + \frac{\rho v}{2D_r} \frac{dv}{d\rho} = D + \frac{\alpha v}{D_r},
\ee 
where $\alpha = (v+\rho \frac{dv}{d\rho})/2$.
Taking Fourier transform of both sides and solving for density mode,
\begin{eqnarray}
\delta \tilde{\rho}(q,\omega) = \frac{-iq}{i\omega + q^2 \mathcal{D}} \tilde{f_d},
\end{eqnarray}
we calculate the dynamic structure factor,
\begin{eqnarray}
S({\bf q},\omega) = \langle \vert \delta \tilde{\rho}(q,\omega)\vert^2 \rangle = \frac{2 V q^2 \Delta_d}{\omega^2 +q^4 \mathcal{D}^2}
\end{eqnarray}
and the static Structure factor,
\begin{eqnarray}
S({\bf q}) = \frac{1}{2\pi} \int_{-\infty}^\infty S(q,\omega)d\omega = \frac{V \Delta_d}{\mathcal{D}},
\end{eqnarray} 
which is independent of ${\bf q}$, i.e., the two-point correlation function $c({\bf r}) \propto \delta({\bf r})$ and correlation length $\xi=0$. This is the reason why the linear analysis is exact in this case and provides the variance exactly
\be
\sigma^2(\rho) = \frac{\Delta_d}{\cal D},
\ee
which is consistent with an integrability condition in \cite{Cates_PRL2008} and with the free energy density function $f(\rho)$ satisfying $d^2f/d\rho^2=1/\sigma^2(\rho)$. This could be seen if we choose $\Delta_d = \tilde D \rho$ as in \cite{Cates_PRL2008} and we find 
\be
\sigma^2(\rho) = \frac{\tilde D \rho}{\tilde D + \frac{\rho v}{2 D_r} \frac{dv}{d\rho}} \simeq \frac{v^2 \rho/2 D_r}{\frac{v^2}{2D_r} + \frac{\rho v}{2 D_r} \frac{dv}{d\rho}} = \left[ \frac{1}{\rho} + \frac{1}{v} \frac{dv}{d\rho} \right]^{-1},
\ee 
by assuming $D \ll v^2/2D_r$. Chemical potential $\mu(\rho)$ can be obtained by integrating fluctuation-response relation (Eq. 5 in the main text) w.r.t. density $\rho$,
\be
\mu(\rho) =\int \frac{1}{\sigma^2(\rho)} d\rho = \ln (\rho v) + c_1,
\ee
$c_1$ an arbitrary constant of integration.

\section*{APPENDIX E: Functional Fokker-Planck Equation and $v=0$ limit}

For a functional Langevin equation (stochastic differential equation) having a general form,
\begin{equation}
\partial_t\rho({\bf r},t)= B[\rho({\bf r},t)] + g({\bf r},t),
\end{equation}
where $B[\rho({\bf r},t)]$ is a functional of $\rho({\bf r})$ and $g({\bf r},t)$ is a Gaussian noise with correlation
\begin{equation}
\langle g({\bf r},t) g({\bf r}^\prime,t) \rangle = G({\bf r},{\bf r}^\prime)\delta(t-t^\prime),
\end{equation}
the functional Fokker-Planck equation is given by \cite{Gardiner}
\bea
\partial_t \mathcal{P}[\rho({\bf r},t)]= -\int d^3 {\bf r} \frac{\delta}{\delta \rho({\bf r})} \lbrace B[\rho({\bf r})] \mathcal{P}[\rho({\bf r},t)] \rbrace
\nonumber \\
 + \frac{1}{2} \int \int d^3 {\bf r} d^3 {\bf r}^\prime \left[ \frac{\delta^2}{\delta \rho({\bf r})\delta \rho({\bf r}^\prime)} G({\bf r},{\bf r}^\prime) \mathcal{P}[\rho({\bf r},t)] \right]
\label{FFP}
\eea

For simplicity, let us consider only one spatial dimension with Cartesian position coordinate $x$. Now, we are interested in a Langevin equation having a particular form
\begin{eqnarray}
\dot{\rho} = - \partial_x \left[ \rho v - {D} (\rho) \partial_x \rho + f_\rho \right]
\end{eqnarray}
where $g(x,t)=\partial_x f_\rho(x,t)$ with noise correlation $\langle f_\rho(x,t) f_\rho(x^\prime,t^\prime) \rangle= 2\Delta(\rho) \delta(x-x^\prime) \delta(t-t^\prime)$. 
Using Eq. \ref{FFP}, the functional Fokker-Planck equation becomes
\begin{eqnarray}
\partial_t \mathcal{P} &=& \int dx \frac{\delta}{\delta \rho(x)} \partial_x \left[\rho v - D(\rho) \partial_x \rho -\Delta(\rho) \partial_x \frac{\delta}{\delta \rho(x)} \right] \mathcal{P}~~~~~~~
\end{eqnarray}
For nonzero $v(\rho) \ne 0$, solution of the above Fokker-Planck equation is not in general known. In a special case, when an integrability condition is satisfied $v/D = \partial_x (\delta F/\delta \rho)$ for a functional $F[\rho(x)]=\int f[\rho(x)] dx$, the steady-state solution is given by the Boltzmann form ${\cal P} \sim \exp[- F[\rho(x)]$ \cite{Cates_PRL2008}.

When velocity $v(\rho) = 0$, i.e., in equilibrium, the Fokker-Planck equation for the many body probability $\mathcal{P}[\rho(x), t]$ can be shown to always have the Boltzmann form as follows. The Fokker-Planck equation in this case can be written as,
\begin{equation}
\dot{\mathcal{P}}=\int dx \frac{\delta}{\delta\rho(x)} \partial_x \left[ -{D}(\rho)\frac{\partial \rho}{\partial x}-\Delta(\rho)\left( \frac{\delta}{\delta \rho} \right)^\prime \right] \mathcal{P}
\label{FPeq}
\end{equation}
We start with an ansatz $\mathcal{P} \sim  \exp\left[-\int f(\rho) dx \right]$ and, using
\begin{eqnarray}
\left( \frac{\delta \mathcal{P}}{\delta \rho} \right)^\prime = -\mathcal{P} \frac{d^2 f}{d \rho^2} \frac{\partial\rho}{\partial x},
\end{eqnarray}
in Eq. \ref{FPeq}, we obtain $f(\rho)$ as given below,
\begin{eqnarray}
&& -{D}(\rho)\frac{\partial \rho}{\partial x} \mathcal{P}-\Delta(\rho)\left( \frac{\delta \mathcal{P}}{\delta \rho} \right)^\prime = 0 \\
\Rightarrow && - {D}(\rho)\frac{\partial \rho}{\partial x} \mathcal{P}+\Delta(\rho) \mathcal{P} \frac{d^2 f}{d \rho^2} \frac{\partial\rho}{\partial x} = 0 \\
\Rightarrow && \frac{\partial \rho}{\partial x} \mathcal{P} \left[-{D}(\rho)+\Delta(\rho) \frac{d^2 f}{d \rho^2} \right] = 0 \\
\Rightarrow && \frac{d^2 f}{d \rho^2} = \frac{{D}(\rho)}{\Delta(\rho)}.
\end{eqnarray}
Therefore the steady-state probability functional for density fluctuation can be written as ${\cal P}[\{\rho({\bf r})\}] \propto \exp[-\int f[\rho({\bf r})] d^2 {\bf r}]$ where $d^2 f/d\rho^2=[\Delta_d(\rho)/D(\rho)]^{-1}$. This is what is expected from the equilibrium fluctuation-dissipation theorem.

\end{document}